# Hysteretic characteristics of a double stripline in the critical state

R M Ainbinder and G M Maksimova

Department of Theoretical Physics, Nizhny Novgorod University, 23 Gagarin Avenue,
603600 Nizhny Novgorod, Russia

**Abstract**
Analytical investigations of the critical state are carried out for a
superconducting stripline consisting of two individual coplanar strips with
an arbitrary distance between them. Two different cases are considered: a
stripline with transport current and strips exposed to a perpendicular
magnetic field. In the second case, the obtained solutions correspond to
'field-like' (for unclosed strips) and 'current-like' (for a long rectangular
superconducting loop) states in an isolated strip to which both a transport
current and a magnetic field are applied with constant ratio.

## 1. Introduction

Many theoretical and experimental works have been devoted to the study of the magnetic characteristics of type-II superconductors of different geometries placed in a magnetic field varying in time [1–12]. The obtained results have led to the conclusion that these characteristics are determined by the type of irreversibility mechanism (bulk pinning or surface/edge barrier) rather than geometrical parameters and their orientation relative to an external magnetic field [7, 13]. On the other hand, a more complex configuration of superconductors brings qualitatively new features of the magnetic-flux distribution. For example, it has been shown that the magnetic behaviour of thin superconducting rings differs significantly from that of discs with the same outer diameter [7, 8]. Another recent study has reported [14] on the critical states of a current-carrying superconducting strip located between the ideal superconducting shields of various geometries. It has been shown that some shielding configurations may lead to substantial transport enhancements in the flux-free region of the strip where the current amplitude may considerably exceed the pinning-mediated critical current.

Quite appealing from both the experimental and application points of view is the double-strip configuration, which represents two parallel coplanar strips. Thus, in [15–17] the phenomenon of magnetic hysteresis, in the basic sensing element of both high-$T_c$ (YBCO) thin-film SQUID magnetometers and Nb thin-film SQUIDs, was studied.

Zhelezina and Maksimova [18] considered the magnetic-flux penetration problem in a superconducting stripline consisting of two pin-free strips placed in the increasing magnetic field. The vortex entry into such a system is controlled by an edge barrier (of the Bean–Livingston or geometric type). In [19], the general solutions were presented for the Meissner-state magnetic field and current-density distributions for a pair of parallel, coplanar superconducting strips carrying arbitrary but subcritical current in a perpendicular magnetic field.

In this paper, we calculate the current density, magnetic field and magnetic moment analytically for a superconducting stripline consisting of two parallel thin-film strips. In section 2 we describe the basic statements of the model. In section 3 we consider the stripline with a transport current. We present an exact solution for wide enough strips arbitrarily separated. The hysteretic loss power calculated in the framework of the Bean model is proportional to the fourth power of a small ac amplitude of the current. In section 4 we study the current and field distribution for two cases. Case (i) describes the behaviour of two unconnected strips in the presence of an applied magnetic field. Case (ii) deals with the critical state, produced when two strips are connected at the ends. We assume that the strips are zero-field cooled, after which the external field is switched on. In section 5 a brief summary is given.

## 2. Basic model

The stripline under consideration consists of two equivalent coplanar strips of width $w$ and thickness $d$, which are infinitely long along the $x$-axis. The distance between the centres of the strips is $2a$ (figure 1(a)). We assume that the strip width $w$ obeys inequality $w \gg \delta = \max(\lambda_\perp, d)$, where $\lambda_\perp = 2\lambda^2/d$ is the two-dimensional screening length and $\lambda$ is the London penetration depth. It would be quite instructive to describe here the main points of our theoretical framework.

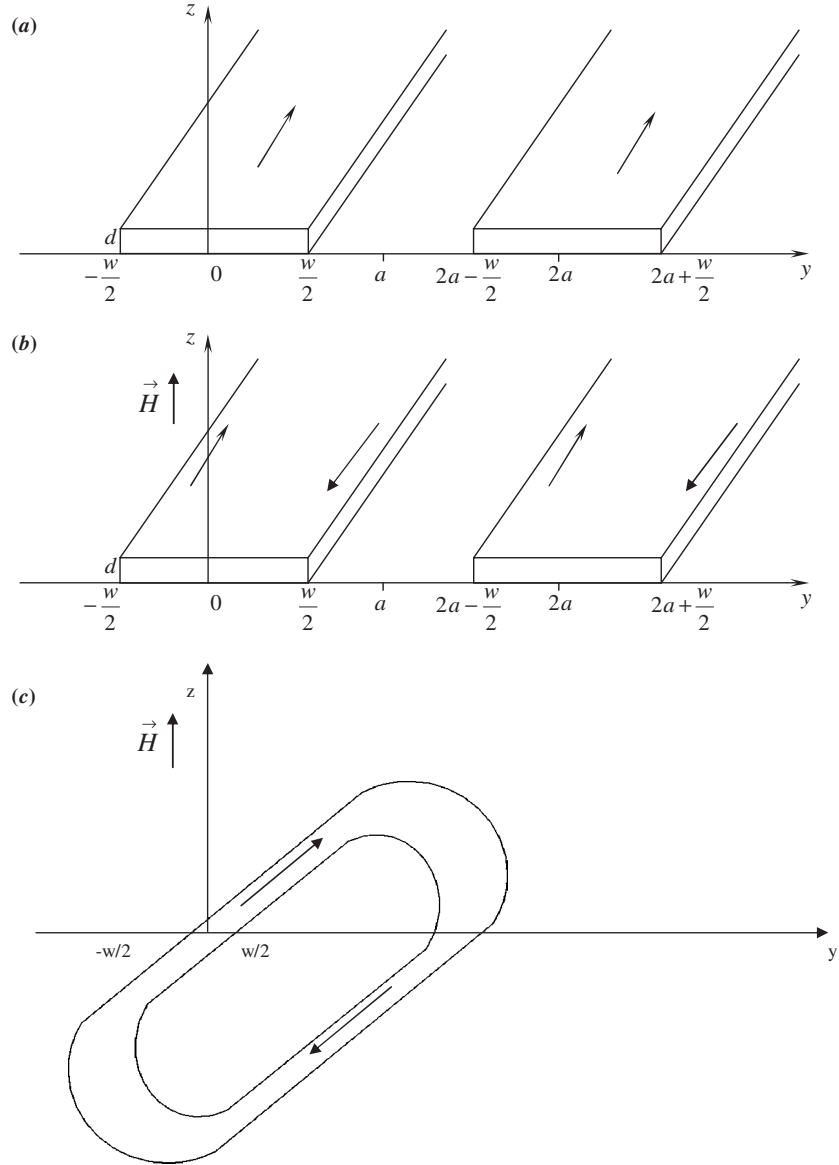

**Figure 1.** Schematic diagram of the film geometries considered in this paper: (*a*) two current-carrying parallel superconducting strips of width $w$, separated by a slot of width $b = (2a - w)$ (see section 3); (*b*) two strips unclosed at their ends in a perpendicular applied magnetic field (see section 4, case (i)); (*c*) double strips closed at their ends (at $x \to \mp\infty$) in a perpendicular magnetic field (see section 4, case (ii)). The arrows indicate the direction of current density.

(1) We calculate the electromagnetic characteristics of the stripline within the Bean model, which assumes a $B$-independent critical current density $j_c$ with the condition that the sheet current is limited to the constant critical value $i_c = j_c d$: $|i(y)| \leqslant i_c$ ($\vec{B}(\vec{r})$ is the local magnetic induction). According to Bean [20], when $j(y)$ locally exceeds $j_c$, the flux lines are rearranged so that condition $|j(y)| \leqslant j_c$ again holds in the entire superconductor. In the case of a long superconductor sample exposed to a parallel field, this assumption leads to a conventional critical state concept, according to which current density is constant ($|j| = j_c$) everywhere in the flux-filled regions of a sample and $j = 0$ in the flux-free zone. However, in the case of a flat superconducting sample exposed to a perpendicular magnetic field, strong demagnetization effects result in a non-zero current density all over the sample, even in the flux-free region. This feature, which is in a visible contrast to the longitudinal geometry, determines the non-trivial magnetic response of thin-film superconductors.

(2) We also restrict our attention to flux densities which are sufficiently high that distinction between $\vec{B}$ and $\vec{H}$ can be ignored, $\vec{B} = \mu_0 \vec{H}$, i.e. disregarding the finite value of the lower critical field $B_{c1} = \mu_0 H_{c1}$ [2–6]. This approximation is quite reasonable when inequality $B > 2 B_{c1}$ holds inside the superconductor [21].

The distribution of the sheet current density $i = \int_0^d j_x(z, y) \, dz$ in the stripline exposed to a perpendicular magnetic field $H$ may be obtained by using the Ampere law

as in [5]

$$\int_{-w/2}^{w/2} \frac{i(y')\,dy'}{y'-y} + \int_{2a-w/2}^{2a+w/2} \frac{i(y')\,dy'}{y'-y} = 2\pi(H - H_i(y)), \quad (1)$$

where $H_i(y)$ is the perpendicular component of the local self-magnetic field connected with the flux-line density $n(y) = \mu_0 H_i(y)/\Phi_0$, and $\Phi_0 = \frac{hc}{2e}$ is the flux quantum.

## 3. Two current-carrying parallel superconducting strips

In this section we describe the behaviour in the absence of an applied field ($H = 0$) when parallel currents $I$ flow in the strips (figure 1(a)). In the absence of a magnetic, the field current distribution is symmetrical with respect to the symmetry axis $y = a$. Therefore, the following relation holds $i(y) = i(2a - y)$, which makes it possible to reduce our consideration to one strip only (for definiteness, the left strip, $-w/2 < y < w/2$). Upon the substitution of a new variable $v = (a - y)^2$ in equation (1), it is reduced to the well-known singular equation of the Cauchy type for $i(v)/\sqrt{v}$

$$\int_{\beta^2}^{\alpha^2} \frac{i(v')\,dv'}{\sqrt{v'}(v'-v)} = -\frac{2\pi H_i(v)}{\sqrt{v}}, \quad (2)$$

where $\alpha = a + w/2$, $\beta = a - w/2$.

The Meissner flux-free state of the stripline implies the absence of a magnetic flux inside, which is equivalent to the disappearance of the perpendicular component of the field at the sheet surfaces: $H_i = 0$. Also, the homogeneous solution to equation (2) corresponding to the total current $2I$ (in both strips) is

$$i(v) = \frac{2I\sqrt{v}}{\pi\sqrt{(\alpha^2 - v)(v - \beta^2)}}, \quad (3)$$

or in the original variables

$$i(y) = \frac{2I(a-y)}{\pi\sqrt{(w^2/4 - y^2)((2a-y)^2 - w^2/4)}}, \quad (4)$$

which agrees with equation (21a) for case C of [19] (parallel currents). Equation (4) holds for all $-w/2 < y < w/2$ (for the left strip) except within the narrow region of width $\delta = \max(\lambda_\perp, d)$ near the edges. We assume hereafter that $w \gg \delta$, so that penetration-depth corrections are needed only in the region of negligible width. It follows from equation (4) that the magnitude of current density at the outer edge ($y = -w/2+\delta$) of the strip exceeds that at inner edge ($y = w/2 - \delta$):

$$\frac{i(y = -w/2 + \delta)}{i(y = w/2 - \delta)} = \sqrt{\frac{a+w/2}{a-w/2}} > 1. \quad (5)$$

When the magnitude of the local self-magnetic field at the edges of the strips exceeds $H_{c1}$, vortices enter the strip and penetrate to a depth determined by the critical-current density $i_c$ and stripline geometries. According to the critical-state model, the current density is then

$$i(v) = \begin{cases} i_c, & \beta^2 \leq v \leq \gamma_2^2, \quad \gamma_1^2 \leq v \leq \alpha^2 \\ \tilde{i}(v), & \gamma_2^2 \leq v \leq \gamma_1^2, \end{cases} \quad (6)$$

and $H_i(v) = 0$ for $\gamma_2^2 \leq v \leq \gamma_1^2$. Substituting equation (6) into equation (2) we find for the function $\tilde{i}(v)/\sqrt{v}$ the integral equation with a singular kernel, which may be inverted to give [22]

$$\frac{\tilde{i}(v)}{\sqrt{v}} = \frac{i_c}{\pi^2\sqrt{(\gamma_1^2 - v)(v - \gamma_2^2)}} \int_{\gamma_2^2}^{\gamma_1^2} \frac{\sqrt{(\gamma_1^2 - t)(t - \gamma_2^2)}}{t - v}\,dt$$
$$\times \left[ \int_{\gamma_1^2}^{\alpha^2} \frac{dx}{\sqrt{x}(x-t)} + \int_{\beta^2}^{\gamma_2^2} \frac{dx}{\sqrt{x}(x-t)} \right]$$
$$+ \frac{C}{\sqrt{(\gamma_1^2 - v)(v - \gamma_2^2)}}. \quad (7)$$

The last term in equation (7) is the general solution of the corresponding homogeneous integral equation. The constants $\gamma_1^2, \gamma_2^2$ and $C$ are determined by the conditions that the current density $\tilde{i}(v)$ should be finite at $v = \gamma_1^2$ and $v = \gamma_2^2$ and the condition that the total current in one strip is $I$. Using these conditions, we obtain from equation (7) the current density $\tilde{i}(v)$ in a field-free region of the strips $\gamma_2^2 \leq v \leq \gamma_1^2$

$$\tilde{i}(v) = \frac{2i_c(\gamma_1^2 - \gamma_2^2)\sqrt{v}}{\pi\gamma_1\sqrt{(\gamma_1^2 - v)(v - \gamma_2^2)}} \left[ \Pi\left(\mu; \frac{v - \gamma_2^2}{\gamma_1^2 - v}; \frac{\gamma_2}{\gamma_1}\right) \right.$$
$$\left. + \Pi\left(\mu; \frac{\gamma_2^2}{\gamma_1^2}\frac{\gamma_1^2 - v}{v - \gamma_2^2}; \frac{\gamma_2}{\gamma_1}\right) - \mathbf{F}\left(\mu; \frac{\gamma_2}{\gamma_1}\right)\right], \quad (8)$$

where

$$\mu = \arcsin\sqrt{\frac{\alpha^2 - \gamma_1^2}{\alpha^2 - \gamma_2^2}} \quad (9)$$

$$\gamma_1\gamma_2 = \alpha\beta, \qquad \frac{i_c}{\alpha}\sqrt{(\alpha^2 - \gamma_1^2)(\alpha^2 - \gamma_2^2)} = I, \quad (10)$$

and $\mathbf{F}(\mu; k), \Pi(\mu; n; k)$ are the incomplete elliptic integrals of the first and third kind, respectively [23]. As follows from equation (10), if $I$ monotonically increases from $I = 0$ to $I_{\max} = i_c w$, $\gamma_1(I)$ decreases continuously from $\gamma_1(0) = \alpha$ to $\gamma_1(I_{\max})$; and $\gamma_2(I)$ increases from $\gamma_2(0) = \beta$ to $\gamma_2(I_{\max}) = \gamma_1(I_{\max}) = \sqrt{\alpha\beta}$. At $I \ll I_{\max}$:

$$\gamma_1^2(I) = \alpha^2\left(1 - \frac{I^2}{i_c^2(\alpha^2 - \beta^2)}\right) \quad (11)$$

$$\gamma_2^2(I) = \beta^2\left(1 + \frac{I^2}{i_c^2(\alpha^2 - \beta^2)}\right). \quad (12)$$

The above solution (6) and (8)–(10) may be tested in the physically transparent situation of direct contact between the sheets, i.e. at $a \to w/2$. In this case $\beta \to 0$, $\gamma_2 \to 0$ so that $\Pi(\mu; n; k) \to \mathbf{F}(\mu; 0)$; thus, two last terms in equation (8) cancel each other. The final result for current distribution at the left side of the film of width $2w$ is

$$i(v) = \begin{cases} \frac{2i_c}{\pi}\arctan\sqrt{\frac{w^2 - \gamma_1^2}{\gamma_1^2 - v}}, & 0 \leq v < \gamma_1^2 \\ i_c, & \gamma_1^2 \leq v < w^2, \end{cases} \quad (13)$$

where $v = (w/2 - y)^2$. At the right side of the film, the current flows symmetrically. This formula coincides with the well-known solution for the thin strip carrying transport current [3, 4, 6]. Note that in these works the method of conformal mapping proposed by Norris [3] was used.

Inserting equations (6) and (8) into equation (2) we obtain the magnetic field component perpendicular to the stripline as a function of variable $\upsilon = (a-y)^2$:

$$H_i\left(\gamma_1^2 \leqslant \upsilon \leqslant \alpha^2\right) = -\frac{i_c\left(\gamma_1^2 - \gamma_2^2\right)\sqrt{\upsilon}}{\pi \gamma_1 \sqrt{(\upsilon - \gamma_1^2)(\upsilon - \gamma_2^2)}}$$

$$\times \left[\Pi\left(\mu; \frac{\upsilon - \gamma_2^2}{\gamma_1^2 - \upsilon}; \frac{\gamma_2}{\gamma_1}\right) + \Pi\left(\mu; \frac{\gamma_2^2}{\gamma_1^2} \frac{\gamma_1^2 - \upsilon}{\upsilon - \gamma_2^2}; \frac{\gamma_2}{\gamma_1}\right)\right.$$

$$\left. - \mathbf{F}\left(\mu; \frac{\gamma_2}{\gamma_1}\right)\right] \quad (14a)$$

$$H_i\left(\gamma_2^2 \leqslant \upsilon \leqslant \gamma_1^2\right) = 0 \quad (14b)$$

$$H_i\left(\beta^2 \leqslant \upsilon \leqslant \gamma_2^2\right) = -H_i\left(\gamma_1^2 \leqslant \upsilon \leqslant \alpha^2\right). \quad (14c)$$

Similar to the case of a single strip, $H_i(\upsilon)$ exhibits a logarithmic singularity at the specimen edges. When $I = I_{max} = i_c w$ the magnetic field for all $\beta^2 < \upsilon < \alpha^2$ is determined by a more simple expression, as follows

$$H_i(\upsilon) = \frac{i_c}{2\pi} \ln\left|\frac{(\alpha^2 - \gamma^2)\sqrt{\upsilon} - \alpha(\upsilon - \gamma^2)}{(\alpha^2 - \gamma^2)\sqrt{\upsilon} + \alpha(\upsilon - \gamma^2)}\right|, \quad (15)$$

where $\gamma = \sqrt{\alpha\beta}$.

So far we have analysed the situation in which the transport current increases monotonically. When the direction of the current is changed, the response of the system depends on the previously attained maximum values. Let $I(t)$ oscillate between the extremum values $\mp I_0$. It suffices to consider one half-period, say, when $I$ decreases from $I_0$ to $-I_0$. The corresponding distribution of the current density and magnetic field has the form [4, 6]:

$$i_\downarrow(y, I, i_c) = i(y, I_0, i_c) - i(y, I_0 - I, 2i_c) \quad (16)$$

$$H_{i\downarrow}(y, I, i_c) = H_i(y, I_0, i_c) - H_i(y, I_0 - I, 2i_c). \quad (17)$$

Figures 2 and 3 show $i(y)$ and $H(y)$ for several values of the transport current $I$. Finally we calculate the hysteretic losses of the stripline, consisting of two coplanar strips when $I(t)$ oscillates with frequency $\nu$ and amplitude $I_0$. It has been shown [3, 4] that the dissipated power $P$ per unit length is uniquely determined by the field profile at peak current. In the considered geometry

$$P = 8\nu i_c \mu_0 \left[\int_{\gamma_1}^{\alpha}(x-\alpha)H_i(x)\,dx - \int_{\beta}^{\gamma_2}(x-\beta)H_i(x)\,dx\right], \quad (18)$$

where $x = a - y = \sqrt{\upsilon}$ and $H_i(\upsilon)$ is given by equations (14a)–(14c) with $I = I_0$. For small and large amplitudes this gives

$$P = \frac{\nu \mu_0}{3\pi}\frac{I_0^4}{I_{max}^2}\left(1 + \frac{w^2}{4a^2}\right), \quad I_0 \ll I_{max}, \quad (19)$$

and

$$P = \frac{4\nu i_c^2 \mu_0}{\pi}\left\{w^2 \ln 2 - w^2 + 2a\left[\left(a + \frac{w}{2}\right)\ln\left(1 + \frac{w}{2a}\right)\right.\right.$$
$$\left.\left. + \left(a - \frac{w}{2}\right)\ln\left(1 - \frac{w}{2a}\right)\right]\right\}, \quad I_0 = I_{max}. \quad (20)$$

We can easily check that, for $a \approx w/2$, equations (19) and (20) coincide with the corresponding formulae $P_{str}(2w, 2I_0)$ for one strip of width $2w$ carrying total current $2I_0$ [3, 4]: $P(a = w/2) = P_{str}(2w, 2I_0)$. The strips become magnetically independent if their spacing is much larger than the width $(a \gg w)$. In this limit natural relation holds $P(a \gg w) = 2P_{str}(w, I_0)$.

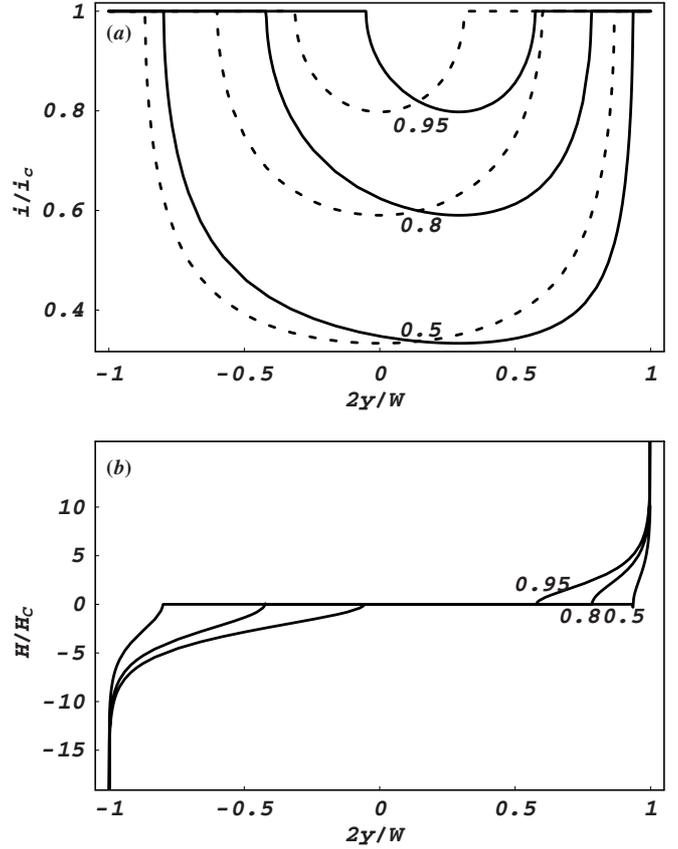

**Figure 2.** Current density $i(y)$ (a) and magnetic field $H(y)$ (b) in a system of superconducting strips of width $w$, separated by a slot of width $b/w = 0.86$, carrying a transport current $I$ which is increased from zero (solid). The depicted profiles are for $I/I_{max} = 0.5, 0.8$ and 0.95. The dashed curves correspond to the case of one strip with transport current.

## 4. Superconducting stripline in a perpendicular magnetic field

The solution found in section 3 for two strips, each carrying transport current $I$, does not depend on whether the strips are closed at $x = \mp\infty$ or not. However, the behaviour of the superconducting stripline in a perpendicular applied magnetic field and $I = 0$ differs in these two geometries.

(i) First we consider the case of two strips unclosed at their ends (figure 1(b)). Using the obvious symmetry $i(y) = -i(2a - y)$ and introducing a new variable $\upsilon = (a-y)^2$ we transpose equation (1) to the equation which coincides with the equation for current density in the isolated strip:

$$\int_{\beta^2}^{\alpha^2}\frac{i(t)\,dt}{t - \upsilon} = -2\pi(H - H_i(\upsilon)). \quad (21)$$

Using the condition $I = 0$, from equation (21) for the Meissner state, we obtain

$$i(\upsilon) = 2H\frac{\left(-\upsilon + \alpha^2 \frac{\mathbf{E}(\rho)}{\mathbf{K}(\rho)}\right)}{\sqrt{(\alpha^2 - \upsilon)(\upsilon - \beta^2)}}, \quad (22)$$

and that the total current in each strip is zero

$$\alpha + \beta = \frac{2\gamma_2^2}{\pi\gamma_1}\left[\sqrt{\frac{\alpha^2-\gamma_1^2}{\alpha^2-\gamma_2^2}}\Pi\left(\frac{\pi}{2};\frac{\xi^2\alpha^2}{\alpha^2-\gamma_2^2};\xi\right)\right.$$
$$\left.+\sqrt{\frac{\gamma_1^2-\beta^2}{\gamma_2^2-\beta^2}}\Pi\left(\frac{\pi}{2};\frac{\xi^2\beta^2}{\beta^2-\gamma_2^2};\xi\right)\right], \quad (26)$$

where $\xi = \sqrt{1-\gamma_2^2/\gamma_1^2}$. As follows from equation (26), the magnetic flux fully penetrates into the strips ($\gamma_1 = \gamma_2$) if $H = \infty$ as for one strip.

The magnetization per unit length of the stripline is defined as

$$m = -2\int_{-w/2}^{w/2} y i(y)\,dy = \int_{\beta^2}^{\alpha^2} i(\upsilon)\,d\upsilon. \quad (27)$$

Substituting equations (23) and (24) into equation (27) we obtain $m$ in an increasing field as

$$m(H, i_c) = -i_c\left(\sqrt{(\alpha^2-\gamma_1^2)(\alpha^2-\gamma_2^2)}\right.$$
$$\left.-\sqrt{(\gamma_1^2-\beta^2)(\gamma_2^2-\beta^2)}\right). \quad (28)$$

When the applied field $H$ oscillates between the extremal values $\mp H_0$ as for the oscillating applied current in section 2, we can show that for $H$ decreasing from $H_0$ to $-H_0$ the current distribution $i_\downarrow$ and magnetic moment $m_\downarrow$ follow from the 'virgin' results (23)–(28):

$$i_\downarrow(\upsilon, H, i_c) = i(\upsilon, H_0, i_c) - i(\upsilon, H_0 - H, 2i_c) \quad (29)$$

$$m_\downarrow(H, i_c) = m(H_0, i_c) - m(H_0 - H, 2i_c). \quad (30)$$

In the half period with increasing $H$ we have

$$i_\uparrow(\upsilon, H, i_c) = -i_\downarrow(\upsilon, -H, i_c) \quad (31)$$

$$m_\uparrow(H, i_c) = -m_\downarrow(-H, i_c). \quad (32)$$

The distribution of the sheet current for different values of magnetic field is plotted in figures 4(a) and 5(a) using equations (23)–(26) and (29). The distance between unclosed strips is fixed at $b/w = 0.86$. Figure 6(a) demonstrates the irreversible magnetization curves $-m(H)$ of two strips unclosed at their ends with aspect ratios $b/w = 0.86, 8$ and $\infty$. It is seen that the hysteretic losses, which are proportional to the area of the hysteresis loop, are reduced when the distance between the strips is reduced.

(ii) Suppose now that a zero-field-cooled superconducting stripline consisting of two parallel strips, connected at their ends (figure 1(c)) is exposed to an increasing applied field. The current and field distribution then at sufficiently low applied magnetic fields may be found from equation (21) taking into account that the magnetic flux into the interior region of the stripline is zero. Specifically, in the Meissner state the sheet current density in the left strip is [18]

$$i(\upsilon) = 2H\frac{(-\upsilon + \alpha^2(1-\mathbf{E}(t)/\mathbf{K}(t)))}{\sqrt{(\alpha^2-\upsilon)(\upsilon-\beta^2)}}, \quad (33)$$

where $t = \beta/\alpha = (a-w/2)/(a+w/2)$. Equation (33) agrees with the results for case E of [19] (zero-flux-quantum state). As follows from equation (33), $i(\upsilon) < 0$ at any ratio of the width of the strips, $w$, and the distance between them, $b$.

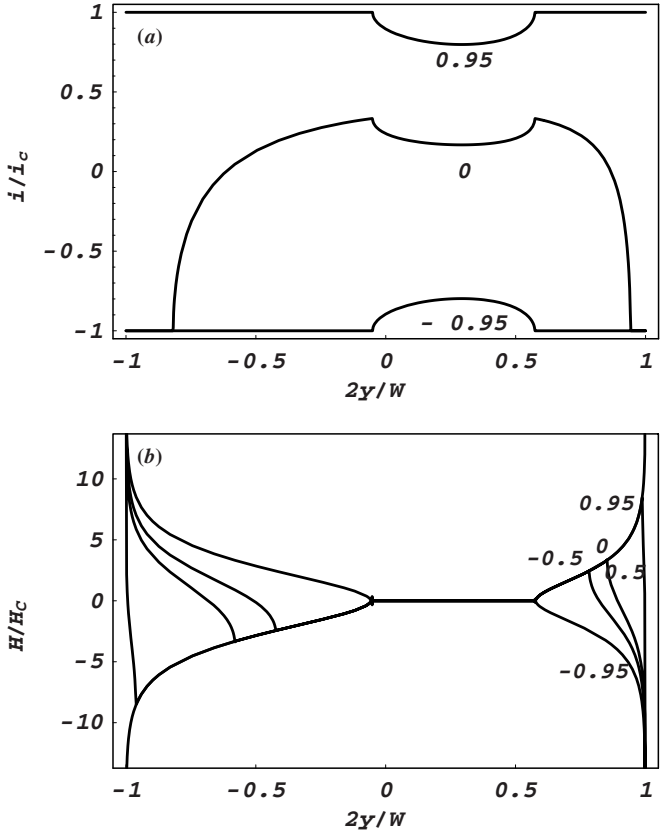

**Figure 3.** Current density $i(y)$ (a) and magnetic field $H(y)$ (b) in superconducting strips when the transport current is reduced from $0.95 I_c$ to $-0.95 I_c$.

where $\rho = \sqrt{1-(\beta/\alpha)^2}$, $\mathbf{E}(\rho)$ and $\mathbf{K}(\rho)$ are complete elliptic integrals of the first and second kind, respectively [23], and $i(\upsilon)$ is the current density in the left strip. Equation (22) agrees with the results for case D of [19] (flux focusing).

In the critical state, using the method described in section 3, we find the solution of equation (21)

$$i(\upsilon) = \begin{cases} i_c, & \beta^2 \leqslant \upsilon \leqslant \gamma_2^2 \\ -\frac{i_c}{\pi}[\arcsin\varphi(\upsilon,\alpha) + \arcsin\varphi(\upsilon,\beta)], \\ & \gamma_2^2 \leqslant \upsilon \leqslant \gamma_1^2 \\ -i_c, & \gamma_1^2 \leqslant \upsilon \leqslant \alpha^2, \end{cases} \quad (23)$$

where

$$\varphi(\upsilon,\alpha) = \frac{(\upsilon-\gamma_1^2)(\alpha^2-\gamma_2^2) + (\upsilon-\gamma_2^2)(\alpha^2-\gamma_1^2)}{(\alpha^2-\upsilon)(\gamma_1^2-\gamma_2^2)}. \quad (24)$$

Note that equations (23) and (24) are the same as for the low-current–high-field regime (or 'field-like' state) for an isolated strip of width $(\alpha^2-\beta^2)$: $\beta^2 \leqslant \upsilon \leqslant \alpha^2$ [6]. The boundaries of the vortex-free region $\gamma_1^2$ and $\gamma_2^2$ (in coordinate $\upsilon$) are determined by the conditions that singular terms of type $\upsilon/\sqrt{(\gamma_1^2-\upsilon)(\upsilon-\gamma_2^2)}$ (which are infinitely large as $\upsilon \to \gamma_1^2, \gamma_2^2$) in the formulae for $i(\upsilon, H, i_c)$ should cancel each other

$$\frac{\sqrt{(\alpha^2-\gamma_1^2)(\gamma_1^2-\beta^2)} + \sqrt{(\alpha^2-\gamma_2^2)(\gamma_2^2-\beta^2)}}{\gamma_1^2-\gamma_2^2} = sh\frac{\pi H}{i_c}, \quad (25)$$

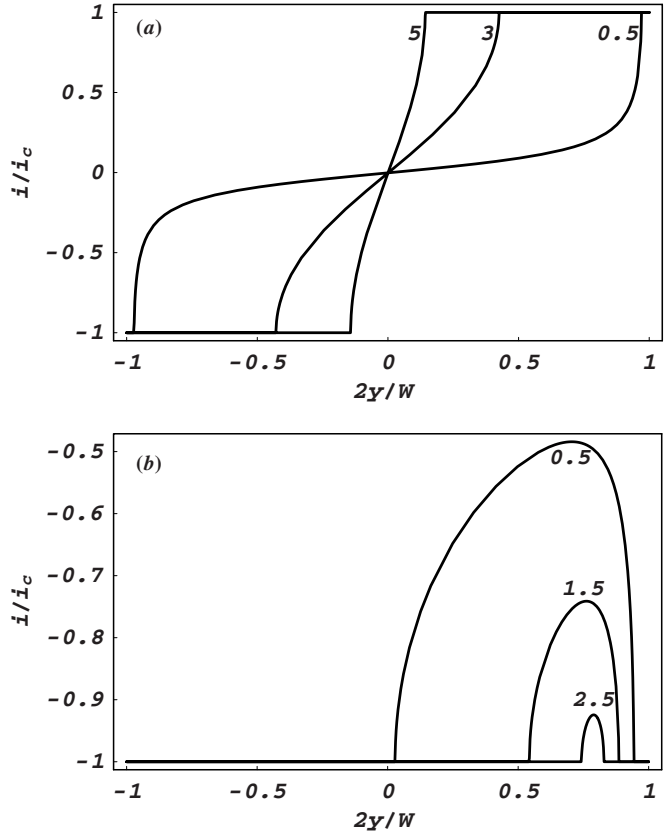
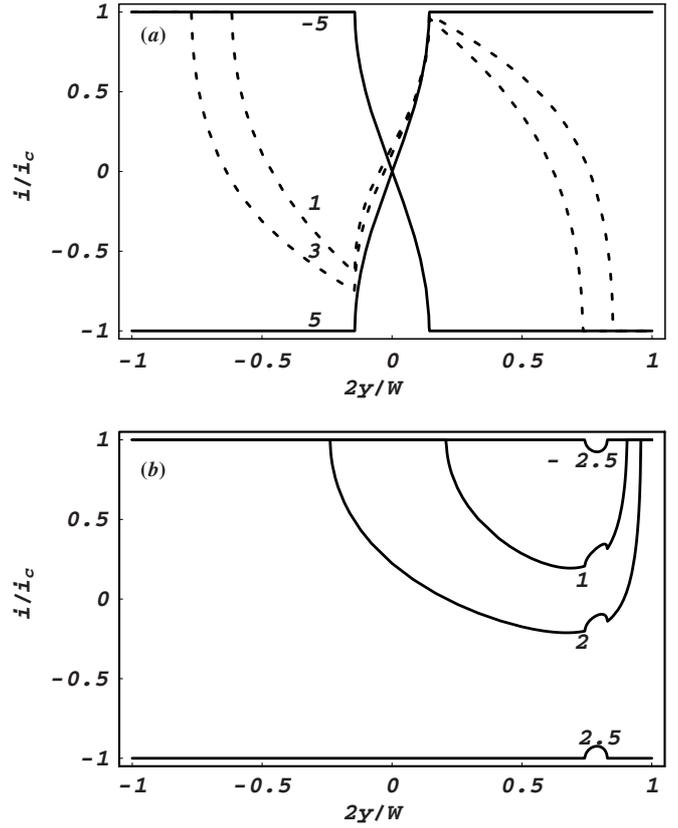

**Figure 4.** Current density $i(y)$ for the double stripline (for the left strip) in a perpendicular magnetic field $H$, which is increased from zero. The depicted profiles are for $h/i_c = 0.5, 3$ and $5$ for unclosed strips (*a*) and for $h/i_c = 0.5, 1.5$ and $2.5$ for double strips connected at their ends (*b*) ($h = 2\pi H$).

**Figure 5.** Current density $i(y)$ for the double stripline as the applied field $H$ is decreased for the strips initially in a partially penetrated state as shown in figure 4.

So, when the self-field at the edges exceeds $H_{c1}$, some positive flux penetrates from the outer edges ($H_i > 0$) and some negative flux penetrates from the inner edges ($H_i < 0$). The corresponding current density is (figure 4(*b*))

$$i(\upsilon) = \begin{cases} -i_c, & \beta^2 \leqslant \upsilon \leqslant \gamma_2^2; \quad \gamma_1^2 \leqslant \upsilon \leqslant \alpha^2 \\ -\dfrac{i_c}{\pi}[\pi + \arcsin\varphi(\upsilon,\beta) - \arcsin\varphi(\upsilon,\alpha)], \\ \gamma_2^2 \leqslant \upsilon \leqslant \gamma_1^2, \end{cases} \quad (34)$$

where $\varphi(\upsilon, \alpha)$ is determined by equation (24). $(\alpha - \gamma_1)$ and $(\gamma_2 - \beta)$ are the flux penetration depths (in original coordinate $y$) at the outer and inner edges of the strips, respectively. The parameters $\gamma_1^2$ and $\gamma_2^2$ are determined by the constraints that (1) the current density should be finite at the boundaries of the vortex-free region (i.e. at $\upsilon = \gamma_1^2$ and $\upsilon = \gamma_2^2$)

$$\frac{\sqrt{(\alpha^2 - \gamma_1^2)(\gamma_1^2 - \beta^2)} - \sqrt{(\alpha^2 - \gamma_2^2)(\gamma_2^2 - \beta^2)}}{\gamma_1^2 - \gamma_2^2} = \sh\frac{\pi H}{i_c}, \quad (35)$$

and (2) that the total magnetic flux inside the interior region of the stripline is zero, $\Phi_{tot} = \Phi_{str}^{in} + \Phi_L = 0$. Here, $\Phi_{str}^{in}$ is the negative magnetic flux, entering the interior region of both strips from the inner edges

$$\Phi_{str}^{in} = \mu_0 \int_{\beta^2}^{\gamma_2^2} \frac{H_i(\upsilon)\,d\upsilon}{\sqrt{\upsilon}}, \quad (36)$$

and $\Phi_L$ is the magnetic flux inside the loop

$$\Phi_L = \mu_0 \int_0^{\beta^2} \frac{H_i(\upsilon)\,d\upsilon}{\sqrt{\upsilon}}. \quad (37)$$

Using equations (21) and (34) to determine $H_i(\upsilon)$ we find from equations (36) and (37) the condition $\Phi_{tot} = 0$:

$$\gamma_2 \ln\frac{\alpha^2 - \gamma_2^2}{\gamma_2^2 - \beta^2} + \alpha \ln\frac{\alpha + \gamma_2}{\alpha - \gamma_2} + \beta \ln\frac{\gamma_2 - \beta}{\gamma_2 + \beta}$$
$$+ \frac{1}{\pi}\int_{\gamma_2}^{\gamma_1}[\arcsin\varphi(x^2,\alpha) - \arcsin\varphi(x^2,\beta)]$$
$$\times \ln\frac{x + \gamma_2}{x - \gamma_2}\,dx = \frac{2\pi H \gamma_2}{i_c}. \quad (38)$$

The magnetic moment per unit length is

$$m(H, i_c) = -i_c\left(\sqrt{(\alpha^2 - \gamma_1^2)(\alpha^2 - \gamma_2^2)} + \sqrt{(\gamma_1^2 - \beta^2)(\gamma_2^2 - \beta^2)}\right). \quad (39)$$

The solution (34) (in combination with equation (24)) found above corresponds (in terms of coordinate $\upsilon$) to the 'current-like' state taking place in a single superconducting strip of width $(\alpha^2 - \beta^2)$ to which both transport current and magnetic field are applied with a constant ratio [6]. But the closed stripline behaves in a qualitatively different way

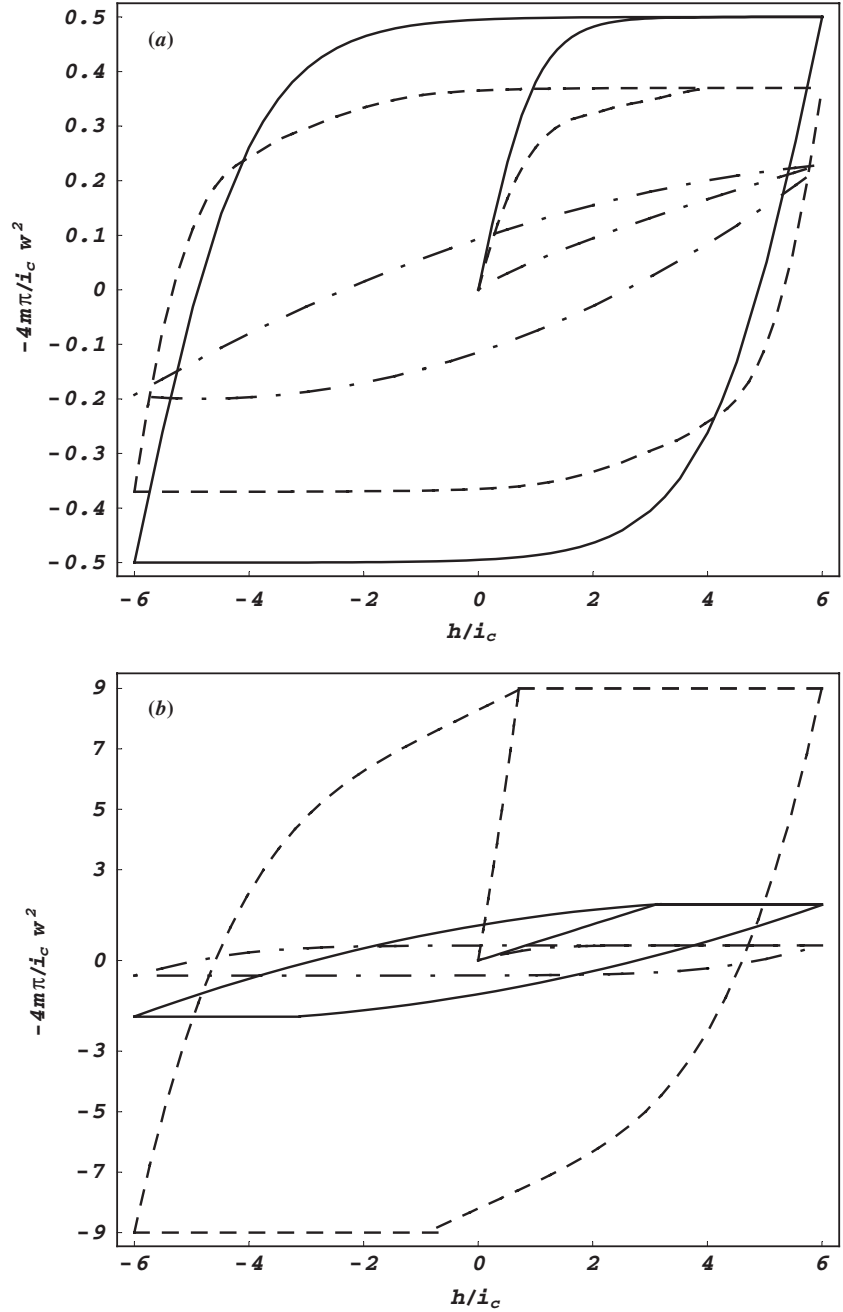

**Figure 6.** The negative magnetic moment $m_{\downarrow\uparrow}$ per unit length of the double stripline in a perpendicular magnetic field $h$ which is cycled with amplitude $h/i_c = 6$ at different values of the slot between them. (a) For two strips unclosed at their ends: the dashed and dash-dotted curves correspond to strips separated by a slot of widths $b/w = 8$ and $0.86$, respectively. The solid curve gives the magnetic moment for one isolated strip of width $w (b/w = \infty)$. (b) The magnetic moment of two superconducting strips closed at their ends and separated by a slot of width $b/w = 8$ (dashed) and $b/w = 0.86$ (solid). The dash-dotted curve corresponds to one strip of width $2w (b/w = 0)$.

compared with both the isolated strip and two unclosed strips (case (i)).

As long as current $I_i$ induced in the stripline by the applied field $H$ is $I_i < I_c$ ($I_c = i_c w$), the magnetic flux into the interior side of the considered circuit remains equal to zero: $\Phi_{tot} = 0$. At $H = H_p$ the field-free region in the strips disappears (i.e. $\gamma_1 = \gamma_2 = \gamma(H_p)$) and the induced current becomes equal to $I_c$. When the applied field is increased further, the vortices of opposite sign move toward each other and annihilate as long as $H$ increases. The parameter $\gamma(H)$, which separates the region containing flux ($\gamma^2(H) < \upsilon < \alpha^2$) and antiflux ($\beta^2 < \upsilon < \gamma^2(H)$), is determined by equation (35) ($\gamma_1 = \gamma_2 = \gamma(H)$):

$$\gamma^2(H) = \beta^2 + \frac{\alpha^2 - \beta^2}{1 + \exp(2\pi H/i_c)}. \quad (40)$$

As follows from equation (40) with the increase of the field $H$ the 'annihilation border' $\gamma^2(H)$ shifts toward the inner edge $\upsilon = \beta^2$. Therefore, the contribution of vortex-related negative magnetic flux $\Phi_{str}^{in}$ into $\Phi_{tot}$ is decreased so that $\Phi_{tot}$ becomes positive at $H > H_p$:

$$\Phi_{tot} = \frac{i_c \mu_0}{\pi}\left[\beta \ln \frac{\gamma(H) + \beta}{\gamma(H) - \beta} - \alpha \ln \frac{\alpha + \gamma(H)}{\alpha - \gamma(H)}\right]. \quad (41)$$

The magnetic moment $m(H, i_c)$ of the stripline at $H > H_p$ is saturated to the value:

$$m_{sat} = m(H_p, i_c) = -2i_c wa. \quad (42)$$

The applied field $H_p$ at which this saturation is reached follows from equations (40) and (41) with $\Phi_{tot} = 0$. It is not difficult to show that if the slot between the strips is very narrow, $a \to w/2$ (the limit of one strip), $H_p$ diverges logarithmically as

$$H_p \approx \frac{i_c}{2\pi} \ln \frac{w^2}{(D^2-1)(a-w/2)^2}, \quad (43)$$

where $D > 1$ satisfies the equation $2D = \ln((D+1)/(D-1))$. This result shows that full penetration of the magnetic flux into the isolated strip becomes possible only at $H \to \infty$. For the stripline with large slot size ($a \gg w$) the corresponding value of $H_p$ is very low

$$H_p \approx \frac{i_c}{2\pi} \frac{w}{a} \left( \ln \frac{4a}{w} + \frac{1}{2} \right). \quad (44)$$

A similar result has been obtained in [7, 8] for a thin narrow ring with a width much smaller than the mean radius. Bean magnetization curves for a closed double stripline with arbitrary parameters may be constructed from the initial slope (39), the saturation moment (42) using expressions (35) and (38) for $\gamma_1^2(H)$ and $\gamma_2^2(H)$ and expressions (40) and (41) which determine the penetration field $H_p$. Equations (30) and (32) are also helpful to generate the full hysteresis loop. Figure 6(*b*) shows the magnetization curve of the superconducting circuit at different values of $b/w$.

## 5. Summary

In this paper we present an analytical solution of the integral equation for the current density and magnetic field distribution for a superconducting stripline consisting of two coplanar wide strips of width $w$ and thickness $d(w \gg \max(d, 2\lambda^2/d))$, which are separated by a slot of arbitrary width $b = (2a - w)$. Constant critical current density $i_c$ is assumed.

These solutions for the stripline carrying an alternating transport current were used for calculating the hysteretic losses $P$. Specifically, it has been shown that for low amplitude of transport current $I_0$ these losses are small: $P \sim I_0^4$ as for one isolated strip [4].

The behaviour of the superconducting stripline in a perpendicular applied magnetic field $H$, with no applied current ($I = 0$), differs greatly in two cases: (i) of unclosed strips and (ii) for a long rectangular superconducting loop. The solution found above for current and magnetic field profiles corresponds to the 'field-like' (i) and 'current-like' (ii) states in an isolated superconducting strip to which both a transport current and a magnetic field are applied with constant ratio [4, 6]. However, the closed stripline behaves differently from the two above cases. Specifically, there exists the penetration field $H_p$ such as that for all $H \geqslant H_p$ the current density in the strips is equal to $i_c$ and the magnetic flux into the interior side of the considered stripline increases from zero with increasing $H$. The magnetic moment of the double strip with connected ends saturates at $H \geqslant H_p$. Note that, as shown in [24], the unintentional presence of a weak spot (weak link or flux leak) in a flat superconducting ring (case (ii)) allows us to observe both the 'current-like' and 'field-like' shielding states and occurs before and after the weak spot becomes transparent to flux motion.

We expect that results obtained in this paper will be useful for studying the electromagnetic response of a superconducting stripline, and specifically for discussing the results obtained by dc magnetization on the superconducting loop in a SQUID magnetometer.

## Acknowledgments

We are grateful to Professor I L Maksimov for his interest in this investigation and for his helpful remarks. The study was supported by the Russian Foundation for Basic Research (project N 01-02-16593), by the Ministry of Science and Technology of Russia (project N 107-1(00)), and by the Ministry of Education of the Russian Federation (project N E-00-3.4-331).